\documentclass{moriond}

\def\be{\begin{equation}}
\def\ee{\end{equation}}
\def\bea{\begin{eqnarray}}
\def\eea{\end{eqnarray}}

\newcommand{\Photo}{\includegraphics[height=35mm]{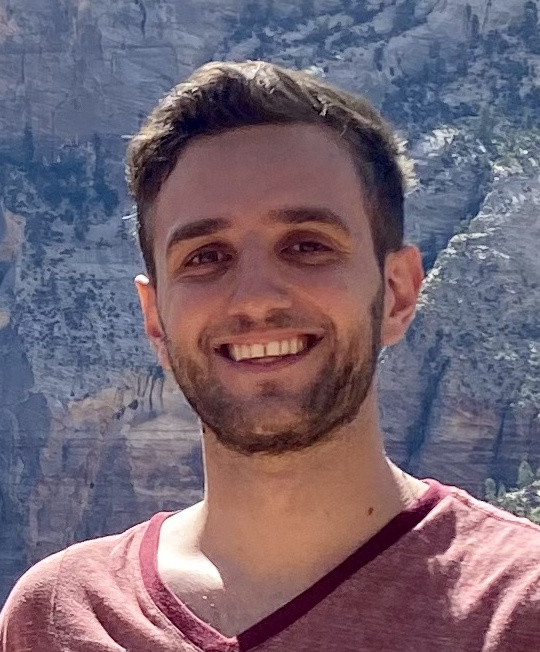}}

\begin{document}
\vspace*{4cm}
\title{Recent measurements of W and Z (+jets) at ATLAS and CMS}

\author{ Oz Amram, on behalf of the ATLAS and CMS Collaborations}

\address{Wilson Hall 11W, Fermilab, Batavia, IL. USA}

\maketitle\abstracts{
  Vector boson measurements at the LHC are an ideal laboratory to study detailed properties of the QCD and electroweak sectors of the Standard Model.
  Several recent measurements of W and Z bosons by the ATLAS and CMS collaborations are presented. 
}

\section{Introduction}
Production of single vector bosons are one of the best understood processes both experimentally and theoretically. 
On the experimental side, such processes have very high cross sections and clean final states with easily identifiable leptons.
Theoretically, they have rich phenomenology  with perturbative QCD and are often a testing ground for higher order calculations
because results can be readily compared to measurements.
Measurements of these processes which focus on details of the production of these bosons can probe various effects in perturbative QCD.
Additionally, measurements of their decay probe the electroweak sector and have sensitivity to potential beyond standard model (BSM) physics.
Over the lifetime of the LHC, many measurements of these processes have been performed by the ATLAS and CMS collaborations.
This contribution highlights several new measurements in this area released this the previous year's Moriond. 
Several relevant measurements were not highlighted in this talk due to time constraints \cite{CMS_Zinv,CMS_AFB,CMS_Zjets}
or being covered in other contributions \cite{ATLAS_Wmass,ATLAS_Zalpha,ATLAS_Z_diff,ATLAS_precision}.

\section{Measurements focusing on QCD effects}

\subsection{W + charm}
Both ATLAS and CMS recently measured the production of a W boson in association with a charm quark \cite{ATLAS_Wcharm,CMS_Wcharm}.
Such measurements are sensitive to the contribution of strange quarks to parton distribution functions (PDF's), which currently have large uncertainties.
The two measurements adopt slightly different approaches. The ATLAS measurement focuses on the production and decay of specific charm mesons : 
$D \to K\pi\pi$ and $D* \to D^0 \pi \to K \pi \pi$.
These meson decays are reconstructed from candidates of 3 displaced tracks.
CMS takes a more inclusive approach and tags an entire jet as charm-like based on secondary vertices or leptons in the jet. 
Both measurements exploit the opposite charge of the W and associated charm meson in order to suppress and estimate backgrounds
by using an "opposite sign minus same-sign" approach.
Comparable precision is achieved by both measurements, and both are found to be consistent with a symmetric strange sea (s=$\bar{\mathrm{s}}$).
The result of the ATLAS measurement is shown in Fig. \ref{fig:fig}.

\subsection{Z $\gamma$ + jets}
ATLAS reported a measurement of the production of a Z boson in association with a photon and jets \cite{ATLAS_Zgamma}.
The measurement focuses on the ISR production of the photon by placing a cut invariant mass of the leptons plus photons. 
The measurement tests the behavior of perturbative QCD by measuring 2D distributions sensitive to Sudakov logarithms of the form
$\alpha^{n}\mathrm{ln}(\frac{p_T}{m})^{n+1}$.
13 different 1D differential cross sections are measured including the $p_T$'s of the jets, Z, various angular quantities and others
Three sets of 2D differential cross sections are measured. Of these 2D distributions one 'resolution' variable is generally being
being probed as a function of a 'hard' variable which is recording the scale.
One such example is $\frac{p_T^{ll\gamma}}{m_{ll\gamma}}$ vs. ${m_{ll\gamma}}$, shown in Fig. \ref{fig:fig}.

\subsection{Z + high $p_T$ jets}
ATLAS reported a measurement of the production of a Z boson in association with a high momentum jet \cite{ATLAS_highpt}.
The measurement probes two distinct angular topologies, one where the Z and the jet are back-to-back and one where 
they are collinear.
Very high $p_T$ ($> 500 $ GeV) and high $S_T$ ($> 600$ GeV) regions are also probed.
Multiple generators were found to describe the data well.

\subsection{Azimuthal correlations in Z + jets}
CMS reported a measurement of correlations between jet multiplicity and angle versus the Z boson $p_T$ \cite{CMS_azimuthal_corr}.
The measurement was done in several different regimes of Z $p_T$: $<10$, $30-50$ and $>100$ GeV.
The measurement demonstrates the transition from "Z + multijet" processes, which are dominant at low Z $p_T$,
to "Z + ISR" which dominates at high Z $p_T$. 
MG5\_aMC@NLO + Pythia8 FxFx with multiple parton interactions was found to agree well with the data.

\subsection{Z + large-R jet}
ATLAS reported a measurement of the production of a Z boson in association with a large radius ($R = 1.0$) jet \cite{ATLAS_largeR}.
The measurement also focused on the case where the large radius jet has been tagged as containing two b-quarks.
This process is an important background for Higgs boson measurements such as $V + H, H\to bb$.
Several 1D distributions sensitive to modeling were measured (jet masses $p_T$'s, and angular separations).
Generators utilizing different formalisms for the inclusion of heavy flavor in PDF's were compared.
The improvement from 5-flavor schemes was evident in the agreement with data.

\section{Measurements focusing on electroweak effects}

\subsection{Search for rare decay of $Z\to\mu\mu\tau\tau$}
CMS reported a first ever search for the rare decay of a Z boson into two muons and two taus \cite{CMS_2tau_2mu}.
In the standard model, this decay is expected to occur with a branching fraction of $\sim 10^{-6}$,
but this could be enhanced by new particles which couple to muons and taus. 
The search focuses on the $\tau \to \mu$ decay modes to suppress backgrounds and lead to a cleanly reconstructed signal. 
The final analysis variable is the invariant mass of the 4 muons, which would be expected to peak slightly below the Z mass due to the 
missing energy in the neutrinos. The observed distribution is shown in Figure \ref{fig:fig}.
No excess above the background prediction is observed.
Limits are set on the ratio of this branching fraction relative to the $Z\to4\mu$ decay at a level approximately 6 times the standard model rate.

\subsection{Measurement of tau polarization in $Z\to\tau\tau$}
\begin{figure}
    \centering
    \includegraphics[width = 0.35 \textwidth]{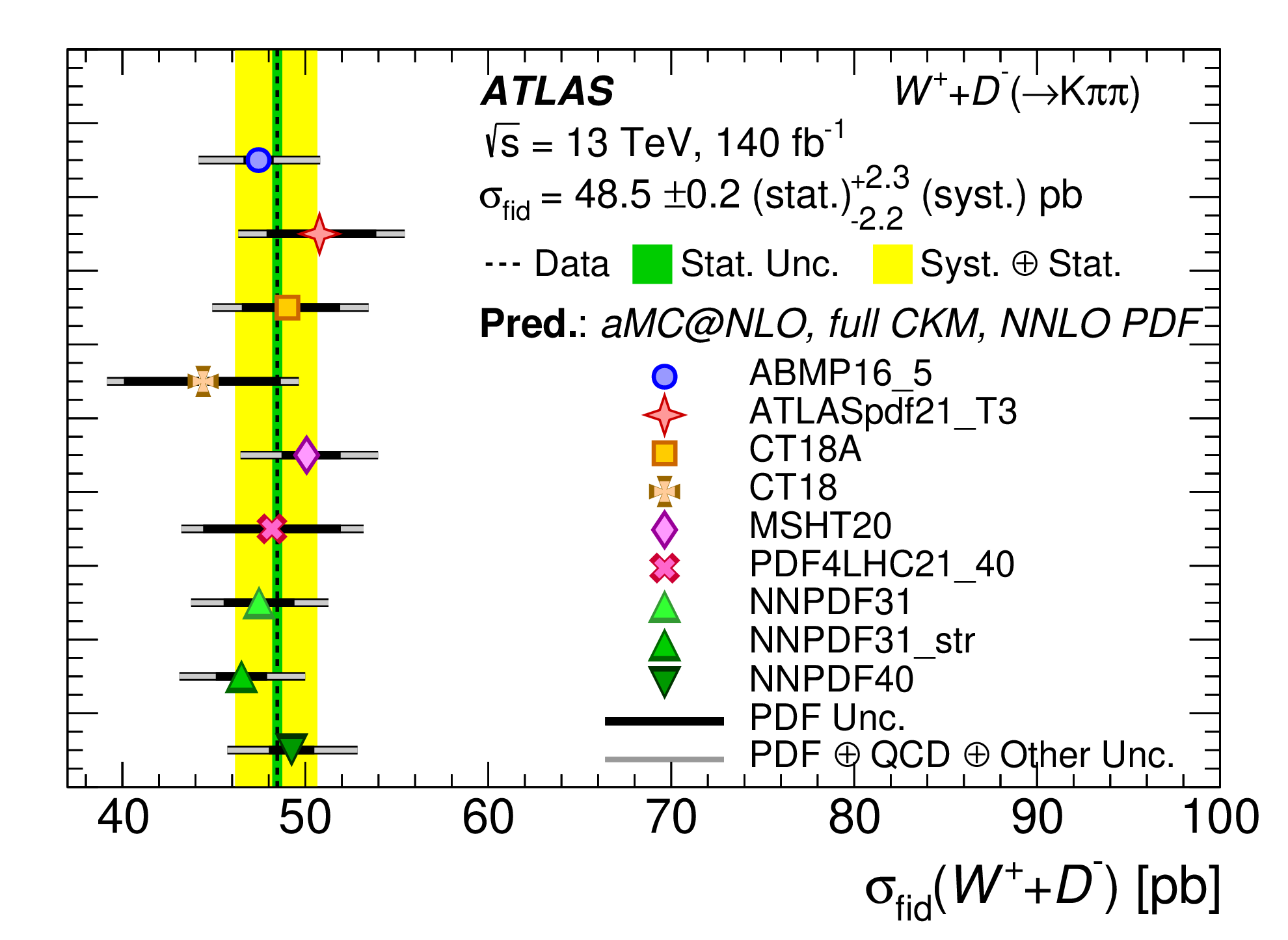}
    \includegraphics[width = 0.35 \textwidth]{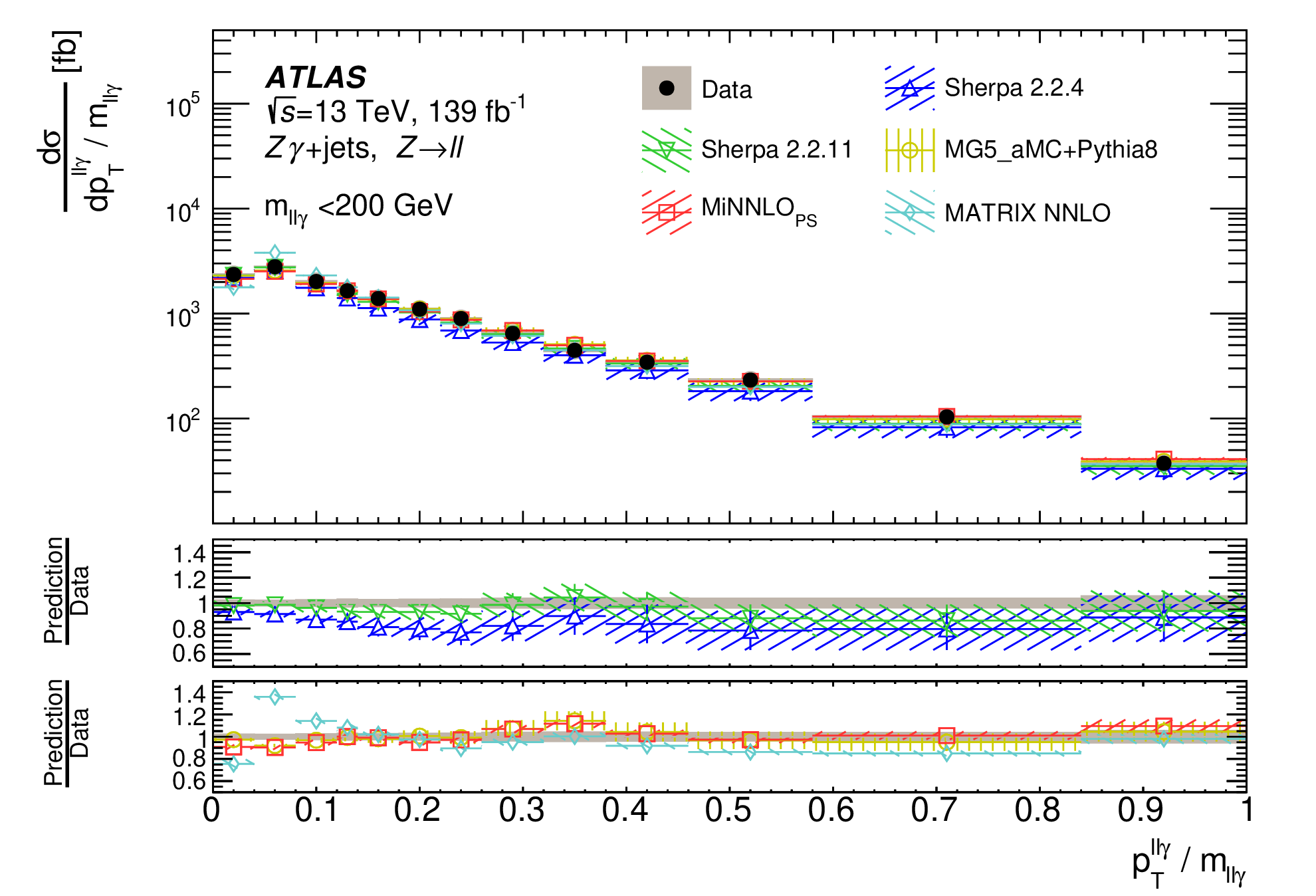}
    \includegraphics[width = 0.35 \textwidth]{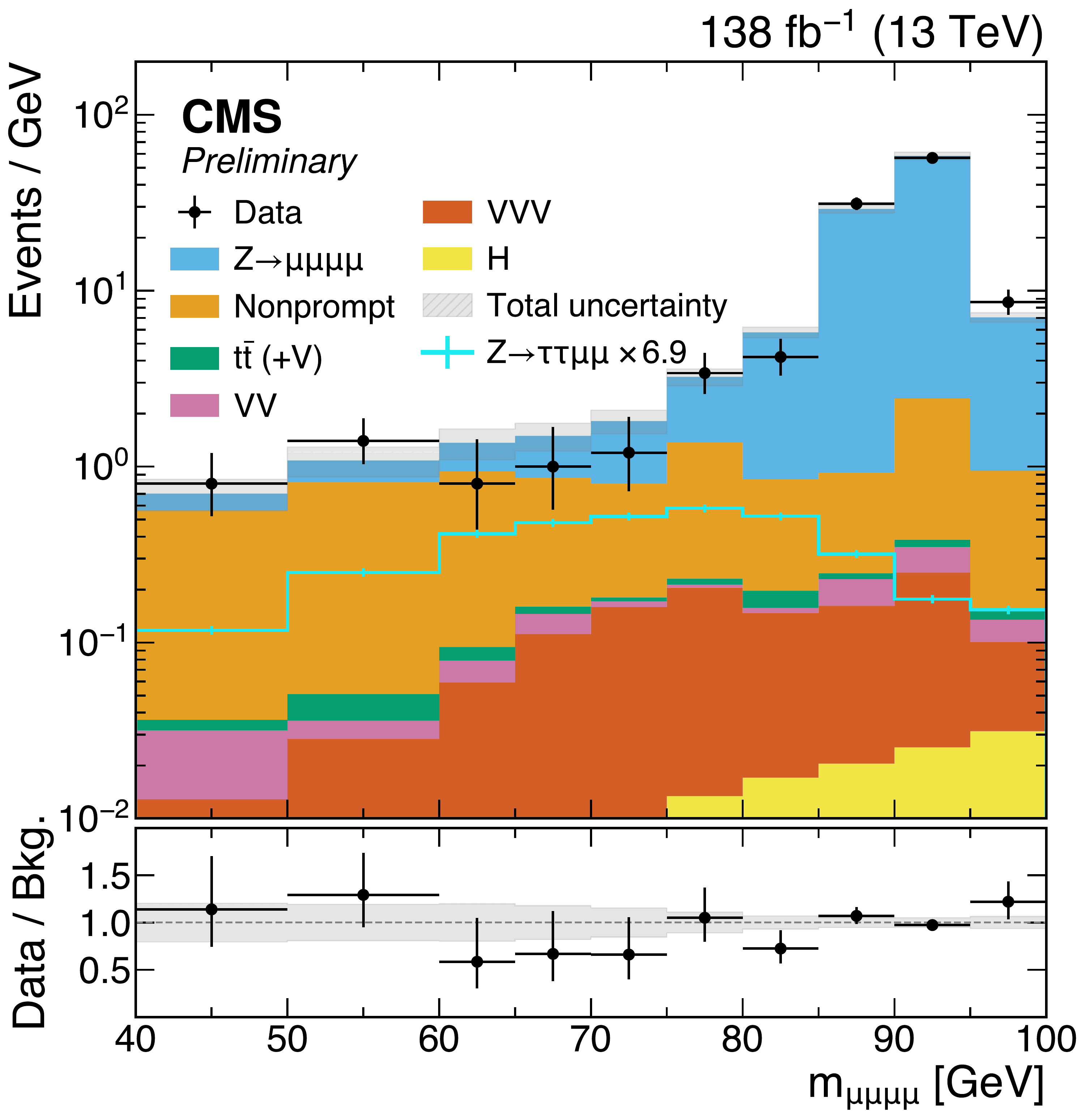}
    \includegraphics[width = 0.35 \textwidth]{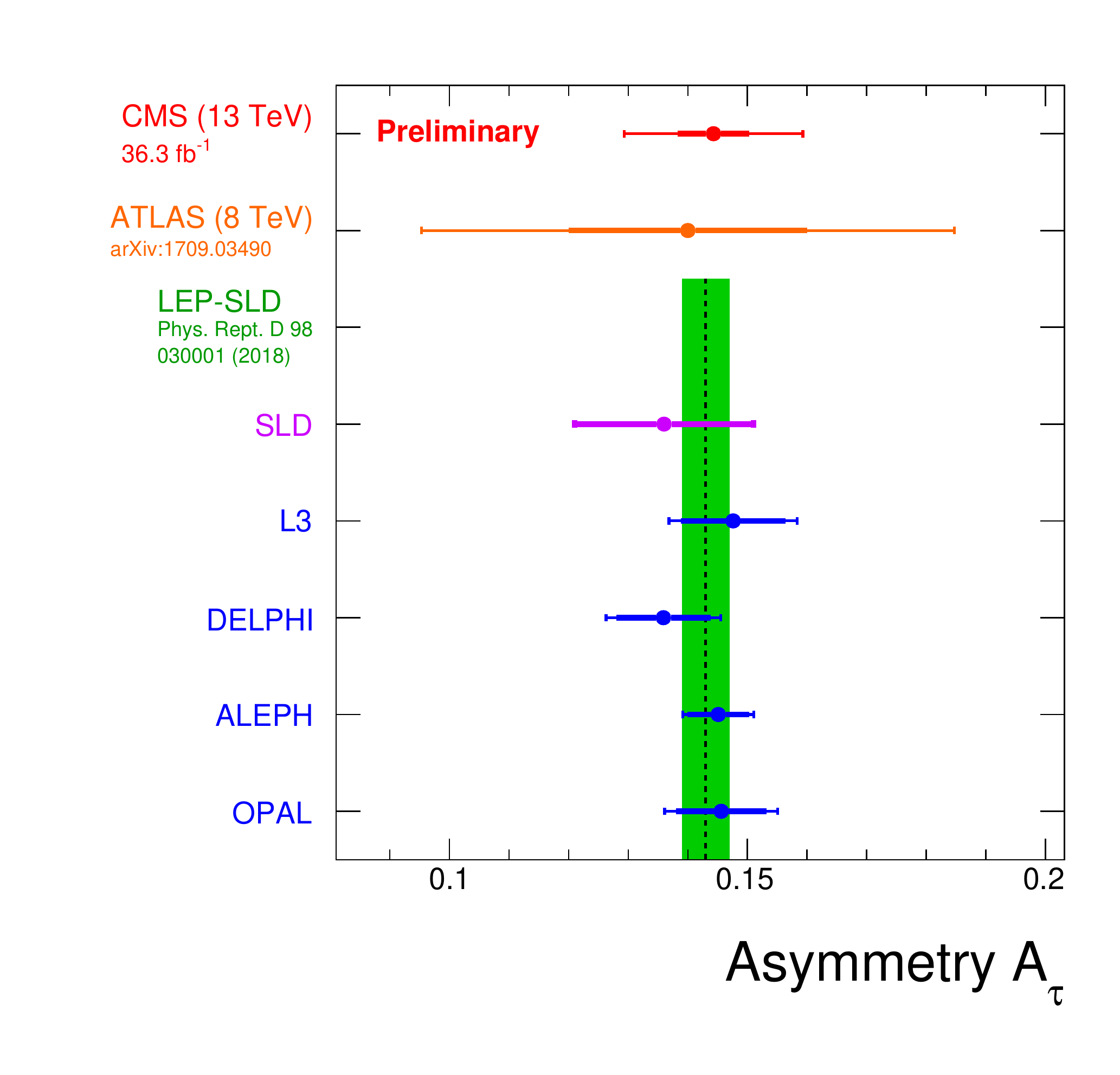}
    \caption{ATLAS measurement of the $W^+ + D^-$ cross section (top left),
      ATLAS measurement of $Z\gamma$ + jets 2D differential distribution (top right),
      CMS search for $Z\to\tau\tau\mu\mu$ (bottom left) and
      CMS measurement of tau polarization (bottom right).
    See text for details.}
    \label{fig:fig}
\end{figure}

CMS reported a measurement of the polarization of $\tau$ leptons in $Z\to\tau\tau$ decays. 
The $\tau$ polarization is sensitive to underlying electroweak parameters $A_{\tau}$ and $\mathrm{sin}^2(\theta_W)$
The measurement considers 11 different categories corresponding to different $\tau$ decay modes. 
Different angular observables are chosen in each category based on the expected sensitivity.
The most sensitive channels were those where one tau decays to a muon and the other decays hadronically.
A simultaneous fit among all 11 categories was performed to extract the polarization.
The result was found to agree with the standard model prediction and achieved a precision comparable to 
single LEP experiments, as shown in Fig. \ref{fig:fig}.

\section{Summary}
ATLAS and CMS presented various results related to the production of single vector bosons.
Measurements related to the production of these bosons will improve our knowledge of QCD, which is crucial as the LHC moves towards the precision era.
Measurements of the electroweak decays of such bosons probe underlying fundamental parameters of the standard model, and have sensitivity to beyond standard model particles.
Both types of measurement will continue to play in important in the physics program of the LHC.

\section*{Acknowledgments}

The author would like to thank the CMS and ATLAS Collaborations for the opportunity to present, and
the organizers of the $57^\mathrm{th}$ Rencontres de Moriond for the successful conference.
This work was partially supported by Fermilab operated by Fermi Research Alliance, LLC under Contract No. DE-AC02-07CH11359 with the United States Department of Energy, and by the National Science Foundation under Cooperative Agreement PHY-2121686.

\section*{References}

\bibliography{WZ_measurements_Amram}

\end{document}